\documentclass[useAMS,usenatbib]{mnras}
\usepackage{times}
\usepackage{url}
\usepackage{graphicx}
\usepackage[usenames]{color}
\DeclareGraphicsExtensions{.pdf,.png,.jpg,.mps,.eps,.ps}
\usepackage{amsmath}
\usepackage{natbib}

\def\unit #1{\,{\rm #1}}

\newcommand\kms{\rm \,\unit{km\,s^{-1}}}

\newcommand\cmsqi{\rm \,\unit{cm^{-2}}}

\newcommand\kev{\rm \,\unit{keV}}

\newcommand\ergs{\rm \,\unit{erg\,s^{-1}}}

\newcommand\funit{\rm \,erg\,cm^{-2}\,s^{-1}}

\newcommand\xiunit{\rm \,erg\,cm\,s^{-1}}

\newcommand\lambdaedd{\lambda_{\rm \, Edd}}

\newcommand\lbol{L_{\rm \, bol}}

\newcommand\msol{M_{\odot}}

\newcommand\nh{\rm N_{H}}

\newcommand\ks{\, \rm ks}

\newcommand\dc{\, \Delta\chi^2}

\newcommand\cd{\,\rm \chi^2/dof}

\newcommand\mpc{\unit{Mpc}}

\newcommand\ev{\unit{\, eV}}

\newcommand\xmm{{\it XMM-Newton}}

\newcommand\nustar{{\it NuSTAR}}

\usepackage{graphicx}

\title{A broadband X-ray spectral study of the Seyfert 1 galaxy ESO~141--G055 with \xmm{} and \nustar{}}

\author [Ghosh et al.] {Ritesh Ghosh$^1$\thanks { Email:ritesh.ghosh1987@gmail.com, riteshghosh.rs@visva-bharati.ac.in} \& Sibasish Laha$^{2,3}$\thanks { Email:sibasish.laha@nasa.gov}, \\ 
$^{1}$Visva-Bharati University, Santiniketan, Bolpur 731235, West Bengal, India.\\
$^{2}$ Astroparticle physics laboratory, NASA Goddard Space Flight Center,Greenbelt, MD 20771, USA.\\
$^{3}$ Center for Research and Exploration in Space Science and Technology (CRESST) and Department of Physics, University of Maryland, Baltimore County,\\ 1000 Hilltop Circle, Baltimore, MD 21250, USA}

\date{\today}
\begin{document}
\pagerange{\pageref{firstpage}--\pageref{lastpage}} \pubyear{2020}




\maketitle
\label{firstpage}

\begin{abstract}

We have extensively studied the broadband X-ray spectra of the source ESO~141--G055 using all available \xmm{} and \nustar{} observations. We detect a
prominent soft excess below $2\kev$, a narrow Fe line and a Compton hump ($>10\kev$). The origin of the soft excess is still debated. We used two models to describe the soft excess: the blurred reflection from the ionized accretion disk and the intrinsic thermal Comptonisation model. We find that both of these models explain the soft excess equally well. We confirm that we do not detect any broad Fe line in the X-ray spectra of this source, although both the physical models prefer a maximally spinning black hole scenario (a$>$0.96). This may mean that either the broad Fe line is absent or blurred beyond detection. 
The Eddington rate of the source is estimated to be $\lambdaedd \sim 0.31$. In the reflection model, the Compton hump has a contribution from both ionized and neutral reflection components. The neutral reflector which simultaneously describes the narrow Fe K$\alpha$ and the Compton hump has a column density of $\rm N_{H} \geq 7\times 10^{24} \rm cm^{-2} $. In addition, we detect a partially covering ionized absorption with ionization parameter $\log \xi/\xiunit $ = $0.1^{+0.1}_{-0.1}$ and column density $\rm N_{H} =20.6^{+1.0}_{-1.0}\times 10^{22} \rm cm^{-2}$ with a covering factor of $0.21^{+0.01}_{-0.01}$.

\end{abstract}


\vspace{0.5cm}

\section{INTRODUCTION}

Active Galactic Nuclei (AGNs) are centres of powerful radiation ranging from radio to Gamma rays. It is commonly believed that a
supermassive black hole (SMBH) accretes matter at a very high rate and thereby emit huge radiation and particle outflows. However,
the SMBHs and the central active engine of an AGN remain spatially unresolved to date in any wavelength band. It is still not clear
how the accretion disk is formed and the different reprocessing media around the SMBH are structured. Broadband X-ray spectra
from AGN, therefore, serves as an important tool to investigate the physics, dynamics and geometry of matter surrounding the SMBH.

The broadband X-ray continua of AGNs are dominated by a powerlaw emission that extends from soft to hard X-rays ($\sim 100 - 500\kev$).
 It is commonly believed to be the result of inverse Compton scattering of the seed UV photons from the accretion
disk by a hot ($\rm T\sim 10^{9}$K) optically thin corona \citep{1973A&A....24..337S, 1993ApJ...413..507H}. Although the exact geometry and
location of the corona is still debated, the powerlaw emission from it has been detected in almost all AGN. The hard X-ray coronal
photons shine onto the accretion disk and other neutral and ionized media around the central engine thereby creating a reprocessed
emission in X-rays, resulting in distinct spectral features. 

Depending on the intensity of X-ray irradiation, the accretion disk becomes ionized and the hard X-ray photons reflected off the 
accretion disk produce siganatures of ionized reflection in soft X-rays (e.g., \citealt{2005MNRAS.358..211R, 2010ApJ...718..695G}) in
the X-ray spectrum. If the reflection happens near enough to the SMBH where it is blurred and distorted by relativistic effects, then
we do not see individual emission lines but a continuous emission in the soft X-rays (e.g., \citealt{1991ApJ...376...90L, 2006MNRAS.365.1067C}).
 In most AGNs, there has been the detection of excess emission in the soft X-rays over and above the powerlaw, popularly known as the soft X-ray
 excess \citep{1985ApJ...297..633S}, the origin of which is still unknown. The blurred ionized reflection is believed to be one of the possible
mechanisms producing it, along with a broad Fe K emission line at $\sim 6.4 - 6.9\kev$. However, other models such as thermal Comptonisation
 has also been proposed for the possible origin of the soft X-ray excess. Despite several theoretical and observational studies
on this matter, the origin of the soft X-ray excess is still debated.

In this work, we take a broadband spectroscopic approach to address the origin of the soft excess, the Fe K line complex and the Compton
 hump at $\rm E> 10\kev$ in the Seyfert 1 galaxy ESO~141--G055 \citep{1978MNRAS.183..129E} where the presence of soft excess and Fe line
 have been affirmatively detected by previous studies \citep{1993ApJ...412...72T, 2003A&A...398..967G}. A recent NuSTAR observation
 in 2016 gives us a rare opportunity to take a hard X-ray ($> 10\kev$) look of this source. The spectral energy distribution is marginally
 variable in both optical/UV and X-ray band \citep{1985ApJ...297..151C, 1992MNRAS.257..677W}, and hence ideal for time-averaged spectroscopic
 studies.

In this paper, we investigate the origin of the soft excess and the hard X-ray Compton hump consistently using broadband spectral 
fitting techniques, in the light of prevalent disk ionized reflection models. The main science questions we address in this work are
 1. Is the disk reflection entirely responsible for both the soft and hard X-ray excess? 2. Do we detect a broad Fe emission line? 
and 3. what fraction of hard X-ray reflection arises from ionized reflection? The paper is organised as follows:
Section \ref{sec:obs} describes the observation and data reduction techniques. The steps taken in the spectral analysis are discussed in
 Section \ref{sec:analysis}. Section \ref{sec:results} discusses the results followed by conclusions in Section \ref{sec:conclusions}.
 Throughout this paper, we assumed a cosmology with $H_{0} = 71\kms \mpc^{-1}, \Omega_{\Lambda} = 0.73$ and $\Omega_{M} = 0.27$.   \\

\section{Observation and data reduction}\label{sec:obs}

ESO~141--G055 was observed five times by \xmm{} in the period from 2001 October 09 to 2007 October 30 and once by \nustar{} in 2016 July 15. The details of the observations and the short notation of the observation ids are listed in Table \ref{Table:obs}. The joint spectral analysis of \xmm{} and \nustar{} provide us with a unique opportunity to study the broad energy range of $0.3-70\kev$ necessary for constraining neutral and disk reflection models. 

We reprocessed the EPIC-pn data from \xmm{} using V18.0.0 of the Science Analysis Software (SAS) with the task {\it epchain} and using the latest calibration database available at that time. We used EPIC-pn data because of its higher signal to noise ratio as compared to MOS. The EPIC-pn data was not available for observation taken in 2001 so we used MOS data instead and reprocessed the data with the task {\it emchain}. Good time intervals are selected from clean calibrated event lists by removing intervals dominated by flaring particle background in energy $>10\kev$, using a rate cutoff of $< 1 \, \rm ct s^{-1}$ for the pn and $< 0.35 \, \rm ct s^{-1}$ for the MOS data respectively. The EVSELECT task was used to select single and double events for both the EPIC-pn ($PATTERN<=4$, $FLAG==0$) and EPIC-MOS ($PATTERN<=12$) source event lists, from a circular source region with a radius of 40 arcsec centred on the centroid of the source image. The background regions were selected with a circle of $40"$ located on the same CCD, but away from the source. The time-averaged source + background and background spectra, as well as the response matrix function (RMF) and auxiliary response function (ARF) for each observation, were calculated using the {\it xmmselect} command in SAS. All the observations were checked for possible pile up using the command {\it epatplot} in SAS and we found that none of them is piled up. The \xmm{} spectra were grouped by a minimum of 200 counts per channel for EPIC-pn and 100 counts per channel for the EPIC-MOS data and a maximum of three resolution elements using the command {\it specgroup}. The optical monitor (OM) camera permits multiwavelength observations of the source simultaneously in the X-ray and optical/UV bands. We reprocessed the OM data using the SAS task {\it omichain} and used only fluxes measured by the UVW2 instrument for consistency as it is the only common filter that is used in all \xmm{} observations. Here we note that the peak wavelength ($2120 \rm \AA$) is relatively nearer to $2500 \rm \AA$ than other filter bands except that for UVM2. The observed UV fluxes were corrected for the Galactic reddening assuming \citet{1999PASP..111...63F} reddening law with $\rm R_{v} = 3.1$ and the Galactic extinction coefficient value used was 0.841.

ESO~141--G055 was observed only once by \nustar{} in 2016 July 15. We reprocessed the \nustar{} data (both FPMA and FPMB) and produced cleaned event files using the standard pipeline NUPIPELINE (V1.9.2), part of HEASOFT V6.27 package, and instrumental responses from NuSTAR CALDB version V20191219. The light curves were created using {\it NUPRODUCTS} command. For light curves and spectra, we used a circular extraction region of $80"$ and $100"$ for source and background respectively. The \nustar{} spectra were grouped by a minimum count of 200 per energy bin, using the command {\it grppha} in the {\it HEASOFT} software.

\section{Data Analysis}\label{sec:analysis}

In our work, we used a set of phenomenological models to fit the broadband spectra to identify the statistically significant spectral features and if they are variable with time. As a next step, we use physically motivated spectral models to obtain the parameter values of interest. This exercise helps to identify the features that are present in the broad-band spectra and can serve as a motivation for the use of specific physical models. We used {\sc XSPEC}~\citep{1996ASPC..101...17A} version 12.10.1, and employed a $\chi^{2}$ minimization technique to fit the data sets. We quote the errors on the best-fit parameters at the $90\%$ confidence level and used Verner cross-sections and Wilms abundances to estimate the effect of Galactic absorption for all the spectral analysis.

\begin{figure}
  \centering 

\hbox{
\includegraphics[width=5.5cm,angle=-90]{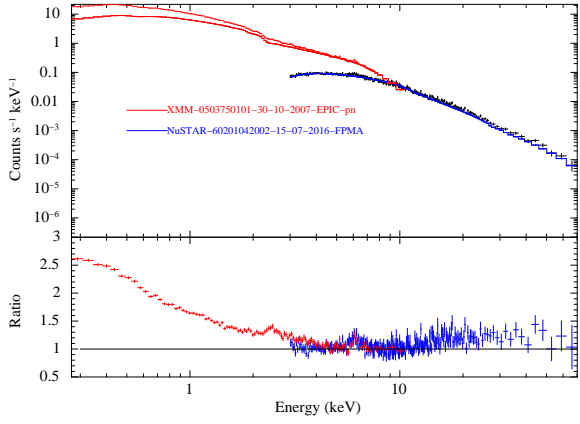}
}\caption{{\it Top panel:} The $0.3-70.0\kev$ joint \nustar{} FPMA (blue) and xmm05 ($80\ks$) EPIC-pn (red) source spectra of ESO~141--G055. {\it Bottom panel:} The $4.0-5.0\kev$ \nustar{} and \xmm{} spectra of ESO~141--G055 fitted with an absorbed powerlaw and the rest of the $0.3-70.0\kev$ dataset extrapolated. The broadband residuals from the fit mentioned above, showing the presence of soft X-ray excess, a Fe emission line, a hard X-ray excess (at $E>10\kev$) and an additional excess at around $2.5\kev$ of unknown origin. The X-axis represents observed frame energy. The X-axis represents observed frame energy. 
} \label{fig:check_excess}

\end{figure}

 The simultaneous spectral fit of \xmm{} and \nustar{} spectra was carried out with a physical set of models. The simultaneous fit helps us to constrain the physical parameters in the broad energy range. Section \ref{subsec:phys} discusses in detail the methods and assumptions followed during these simultaneous spectral fits. A preliminary look at the spectra of the source by fitting an absorbed  powerlaw model in the $4-5\kev$ energy band and extrapolating to the rest of the X-ray band reveals a prominent soft excess (at $\rm E<2.0\kev$) for all \xmm{} observations along with a Fe line complex at $6-7\kev$ whereas, the \nustar{} data revealed a hard X-ray excess above $>10\kev$ (See Fig.~\ref{fig:check_excess}). The value of Galactic absorption column density ($4.9\times 10^{20}\cmsqi$) is taken from \citet{1990ARA&A..28..215D}. An additional emission line of unknown origin is detected at $2.5\kev$ for all the \xmm{} observations.

\subsection{The phenomenological models}\label{subsec:pheno}

\begin{figure}
  \centering 

\includegraphics[width=6.0cm,angle=-90]{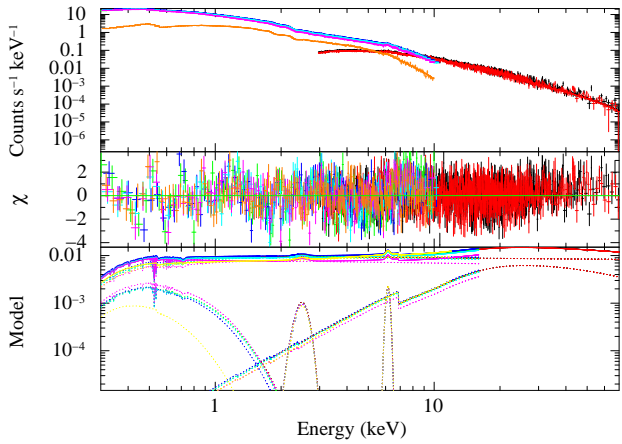}
\caption{ The $0.3-70.0\kev$ \nustar{} and \xmm{} spectra of the source ESO~141--G055 with the best-fitting phenomenological model and residuals. The X-axis represents the observed frame energy. } \label{fig:pheno}

\end{figure}

The baseline phenomenological model used here consists of a neutral Galactic absorption (tbabs), a photoionized absorption model for partially ionized material (zxipcf), a multiple blackbody component to model the soft X-ray excess \citep[diskbb,][]{1984PASJ...36..741M}, the coronal emission described by a power law and the {\it PEXRAV} model to account for the reflection component from the neutral medium. We used two Gaussian profile to describe the excess emission at around $6.4\kev$ and $2.5\kev$. An energy-independent multiplicative factor was used to account for the relative normalization between different instruments of \xmm{} and \nustar{}. In XSPEC notation, the phenomenological model reads as {\tt constant$\times$ tbabs$\times$ zxipcf$\times$(powerlaw+diskbb+zgauss+pexrav)}. 
The inclusion of a partially ionized absorption model {\it zxipcf}, that uses a grid of XSTAR \citep{2001ApJS..133..221K} photoionized absorption models (calculated assuming a microturbulent velocity of $200 \kms$) significantly improved the statistics for all \xmm{} observations. We did not detect intrinsic neutral absorption excess of the Galactic absorption for any of the \xmm{} observations. The \xmm{} energy band ($<10\kev$) is in-sufficient to constrain the {\it pexrav} reflection parameters. Upon using \nustar{} there is a statistically significant improvement (See Table~\ref{Table:pheno}). The phenomenological model provides a satisfactory description to the \nustar{} as well as \xmm{} spectra (see Fig.~\ref{fig:pheno}). We note that an additional excess in the residual at $0.5\kev$ was present for xmm04 and xmm05. Addition of a Gaussian line profile to the set of phenomenological models yielded a much better fit for both these observations ($\cd =171/134$ and $201/147$ for xmm04 and xmm05 respectively). The best-fit parameters and the fit statistics obtained using the phenomenological models are quoted in Table \ref{Table:pheno}. We also included the improvement in statistics ($\rm \dc/dof$) for each spectrum on the addition of the discrete emission line models to determine the statistical significance of the model.

The Fe $\rm K_{\alpha}$ line detected at $6.4\kev$ in the \nustar{} observation was fitted with a Gaussian which yielded an equivalent width of $98\pm 21 \ev$. The improvement in the fit statistics is $\rm \dc/dof$ = 113/3, which reflects a $> 99.99$ per cent confidence in the detection of the emission line \citep{1976ApJ...208..177L}. For \xmm{} observations, a similar approach provided an equivalent-width values of $76 \pm 12 \ev$, $52 \pm 11 \ev$, $55 \pm 17 \ev$, $56 \pm 19 \ev$ and $73 \pm 19 \ev$, for xmm01, xmm02, xmm03, xmm04 and xmm05 respectively with significant improvement in the fit statistics (see Table~\ref{Table:pheno}). Next, we replaced the Gaussian profile with the {\it diskline} model \citep{1989MNRAS.238..729F} to check the presence of broad Fe emission in the spectra. We did not find any significant improvement in the fit statistics ($\rm \dc/dof = 7/3 $) for the \nustar{} data. The fit worsens for \xmm{} observations ($\rm \dc =$ 5, 9, 3, 6, and 7 for xmm01, xmm02, xmm03, xmm04, and xmm05, respectively) and we could not constrain model parameters e.g., the inner radius, the emissivity index and the inclination angle. These results indicate that a broad emission line is not required for any of these observations. 

\subsection{The physical models}\label{subsec:phys}
 
 The use of phenomenological models has confirmed the presence of soft and hard excess in the broad X-ray energy band. In this section, we try to simultaneously describe both soft and hard X-ray excess using two sets of physical models. The first model assumes that both the soft and the hard excess arise from a disk ionized reflection. In XSPEC notation, the first set of models reads as {\it(constant $\times$ tbabs$\times $zxipcf$\times$(relxill + MYTorus))}. Here, {\it relxill} model \citep[version 1.3.5,][]{2014ApJ...782...76G} describes the soft X-ray excess, the power-law continuum and the reflection of primary
hard X-ray photons off an ionized accretion disk while {\it MYTorus} \citep{2009MNRAS.397.1549M,2012MNRAS.423.3360Y} models the cold, distant, neutral reflection from Compton thick medium that contributes to the Compton hump. The photon index $\Gamma$ in {\it MYTorus} and the primary power-law component of the {\it relxill} model were tied together. The {\it MYTorus} inclination angle, the equatorial column density and the normalization of the individual components, i.e., the {\it MYTorus line} and the {\it MYTorus scattered}, were left free to vary. The inclination angles of the {\it relxill} and {\it MYTorus} models were not tied as they are two entirely different reprocessing medium and were treated as separate quantities. As with phenomenological models, we have added a partially covering ionized absorption model {\it zxipcf} for all the observations, to account for any ionized absorption along the line of sight. We checked for possible variation in the ionization parameter and the covering fraction of the partial absorber. The best-fit parameter values are within $3\sigma$ errors for all the observations and hence tied with each other. For all the observations, a Gaussian line profile is added to fit the emission at $2.5\kev$ whose origin is unknown. The {\it relxill} model assumes a coronal or lamp-post geometry of scattering, where the source of primary X-ray emission (corona) irradiates the accretion disk from the top and the hard X-ray photons get Compton scattered from the accretion disk. Hence, it accounts for the soft X-ray excess and the relativistic hard X-ray emission above $10\kev$, contributing to the Compton hump. A broken power-law emissivity profile ($\rm E(r)\propto \rm r^{-q}$), is typically assumed by the {\it relxill} model where E is the emissivity of the gas due to reflection and q is the emissivity index. In Newtonian geometry, for a point source, the emissivity at a large radius out from the source has a form $\rm r^{-3}$. In the general relativistic high-gravity regime the inner emissivity profile will depend upon the location, spatial extent and geometry of the source and calculations \citep{2001MNRAS.321..605D, 2003MNRAS.344L..22M, 2011MNRAS.414.1269W} suggest a very steeply falling profile in the inner regions of the disk. Recent studies of accretion disk/corona emission using the {\it relxill} model support this scenario \citep{2018MNRAS.479.2464G, 2018ApJ...855....3T}. The {\it relxill} model parameter $\rm r_{br}$ determines this transition from relativistic geometry to Newtonian geometry. Thus in our fits, the emissivity index of the inner part of the accretion disk at $\rm r < \rm r_{br}$ has values ranging from $3 < q < 10$, while at the outer part of the disk at $\rm r > \rm r_{br}$, the index is fixed to 3.

The best-fitting model parameters along with the best-fitting statistics for \nustar{} and \xmm{} observations are listed in Table \ref{Table:relxill}. The model parameters that are unlikely to change within the human timescale were tied to each other, e.g., the black hole spin, the inclination angle, the ionization parameter of the accretion disk. The photon index $\Gamma$, the normalization flux and the reflection fraction of all the observations were made free to vary. For the \xmm{} observations, the {\it MYTorus} parameters are not well constrained possibly due to absence of the $> 10\kev$ spectrum and were tied with \nustar{} best-fit parameter values. 
The simultaneous analysis of all the observations with this set of physical models produced a good fit with the fit statistic $\cd = 2314/1997$. We esimated the high energy cutoff of the primary power-law component using the \nustar{} observation. We found a lower limit on the high energy cutoff ($\rm E_{c}$) to a value of $\rm E_{c} >276\kev$. We tied this value for all other observations. We also tied the inner radius for all \xmm{} observations for better constrain on the black hole spin. The broadband spectra infer the presence of a maximally spinning black hole ($a> 0.96$). The fit statistic becomes significantly poor ($\cd = 2713/1998$) when the spin parameter is frozen to zero. We have carried out two separate fits in order to ascertain if the data prefer a maximally spinning black hole. In the first case, we kept the inner radius of the relativistic reflection model fixed to the inner circular stable orbit for a non-rotating black hole ($\rm r_{in} = 6\rm r_{g}$). In the second case, we fixed the inner radius to that of a maximally spinning black hole ($\rm r_{in} = 1.23\rm r_{g}$ ). We note that the broadband data prefer the maximally rotating scenario as the fit statistically worsens by $\dc \sim 300$ for the non-rotating one. We get a poor ($\cd = 2637/2000$) fit if we remove the partially covering ionized absorber model, implying that the model is statistically required by the data. 

Two other flavours of {\it relxill} model, {\it relxillD} and {\it relxilllP} have also been used to test out different reprocessing geometries. The model {\it relxillD} is same as {\it relxill} but allows a higher density for the accretion disk (between $\log N/\rm cm^{-3} =15$ to $\log N/\rm cm^{-3} =19 $). The {\it relxilllP}, on the other hand, assumes a lamp post geometry and the disk is split into multiple zones, which see a different incident spectrum due to relativistic energy shifts of the primary continuum. The use of {\it relxillD} and {\it relxilllP} provided relatively poor fit compared to the {\it relxill} model. For {\it relxillD}, best-fit statistics is $\cd = 2422/1993$ and for {\it relxillP}, $\cd = 2435/1996$. The change in accretion density is not significant and found to be $\log N/\rm cm^{-3} = 15.3^{+0.4}_{-0.5}$ for \nustar{} and $\log N/\rm cm^{-3} = 16.4^{+0.6}_{-0.8}$ for \xmm{} observations respectively. Interestingly, no significant variation is noticed between onservations in the height of the X-ray emitting source $\rm h \sim 3\rm r_{g}$.

The second set of models we used assumes that the soft X-ray excess arises from thermal Comptonization of disk photons by a warm ($\rm T\sim 0.5-1\kev$) and optically thick ($\tau\sim 10-20$) corona \citep{1998MNRAS.301..179M, 2012mnras.420.1848d}. The hard excess, in this case, arises exclusively due to neutral Compton reflection. The broadband model in XSPEC notation reads as {\it(constant $\times$ tbabs$\times $zxipcf$\times$(optxagnf + MYTorus))}. In the model {\it optxagnf} \citep{2012mnras.420.1848d}, the gravitational energy released in the accretion process serves as the power source of the disk emission in UV, the soft X-ray excess, and the power-law emission. The normalization flux of this model is determined by the black hole mass, the Eddington rate, the black hole spin and the distance. Hence the model normalization is frozen at unity. The SMBH mass used $\sim 3.98\times 10^{7}\msol$ was obtained from \citet{2016MNRAS.458.2454L}, who derived it from the average values of stellar velocity dispersion. Similar to the ionized reflection model, we followed a simultaneous fitting approach with all the data sets. The black hole spin, the optical depth of the soft Comptonisation component and the coronal radius were tied between the observations. The Eddington ratio, the photon index ad the parameter $\rm f_{pl}$, that determines the fraction of the power below the coronal radius emitted in the hard comptonisation component, were made free to vary. The fit resulted in comparable statistics ($\cd = 2370/1997$) as that of our first set of physical models however,  we noticed some excess in the residual at $\sim 0.5\kev$. Addition of a Gaussian line profile to this feature provided an improved fit ($\cd = 2335/1994$). The best-fit line energy, $0.55^{+0.02}_{-0.02}$ indicate this excess to be an OVII emission line. Table \ref{Table:optxagnf} lists the best-fitting parameters obtained using this model. A detailed discussion of these results is followed in the next section.

\section{Results and Discussion}\label{sec:results}

\begin{figure*}
  \centering 

\includegraphics[width=8.5cm,height=6.5cm,angle=0]{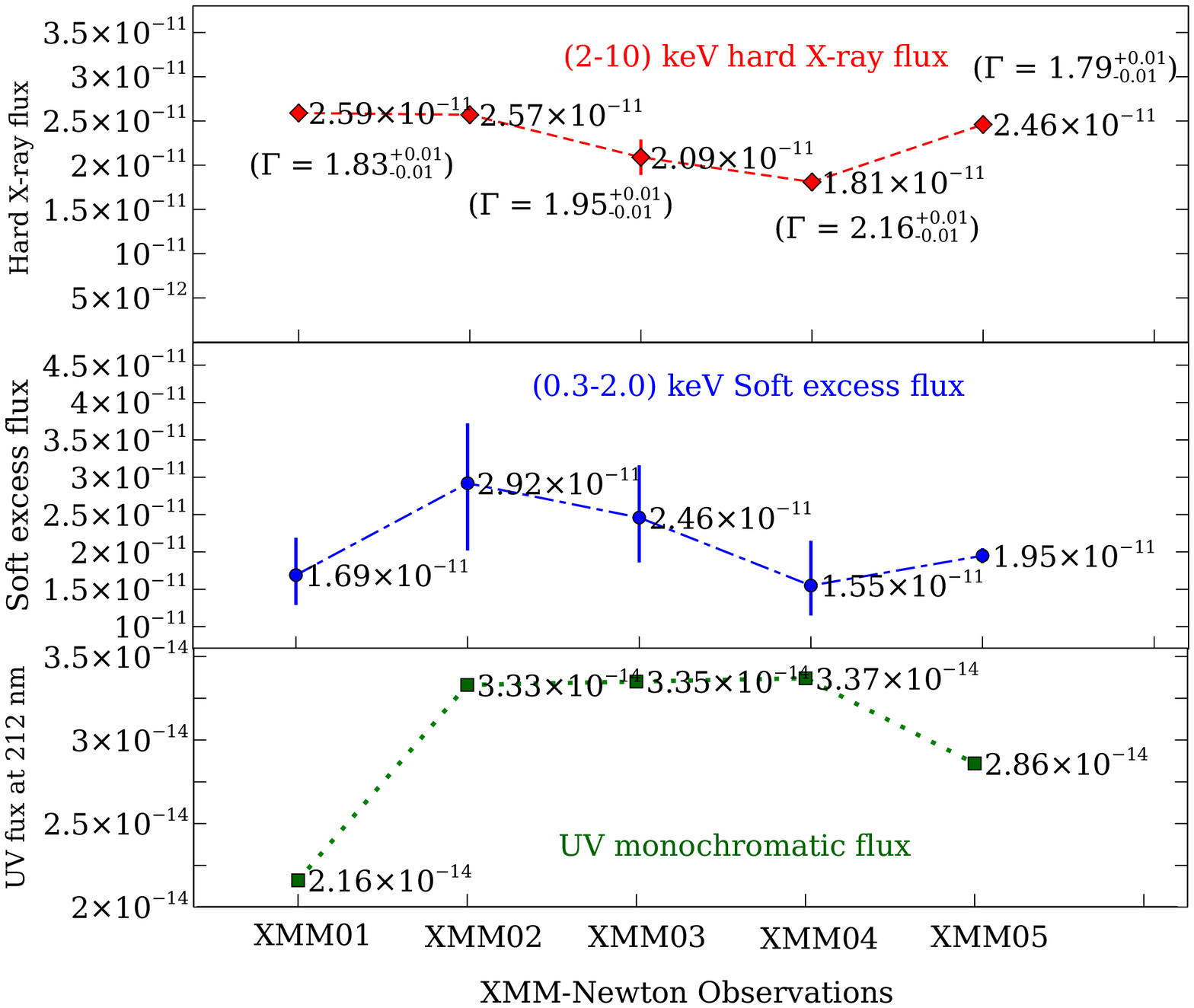}
\caption{ The soft excess (in $0.3-2.0\kev$), the hard ($2.0-10.0\kev$) X-ray and the UV monochromatic flux (at UVW2; 212 nm) variation of five \xmm{} observations of ESO~141--G055 are plotted. The measured soft excess, the $2-10\kev$ and the UV monochromatic (212nm) flux does not show any significant change between observations and are shown in circles (blue), diamonds (red) and squares (green) respectively. The X-axis represents the \xmm{} observations and the Y-axis represents the measured flux in units of $\funit$. 
} \label{fig:variability}

\end{figure*}

We have investigated the broadband X-ray spectra of the Seyfert 1 galaxy ESO~141--G055 using the latest observations from \nustar{} and \xmm{} telescopes.  
In our work, for the first time, we analysed all the \xmm{} and \nustar{} observations covering a broad X-ray energy band ($0.3-70\kev$). 
Using physically motivated models, we carried out a detailed study investigating the presence and origin of the observed reflection features in the source spectra. In our analysis, we find that both ionized disk reflection and intrinsic thermal Comptonisation model yielded equally acceptable statistical fits. The soft X-ray excess ($0.3-2.0\kev$), the hard ($2.0-10.0\kev$) X-ray flux and the UV monochromatic flux at UVW2 (212 nm) for the \xmm{} observations does not show any significant change between observations (See Fig.~\ref{fig:variability}). The $2-10\kev$ unabsorbed luminosity ($ \log \rm L_{2-10\kev}$) of ESO~141--G055 is consistent between 2001 and 2016. The observed values are $43.92\ergs$, $43.99\ergs$, $43.95\ergs$, $43.87\ergs$, $43.96\ergs$ and  $43.93\ergs$ for xmm01, xmm02, xmm03, xmm04, xmm05 and \nustar{} observation respectively. Following the equation, 
$\log \kappa_{\rm Lbol}=1.561-1.853\times \alpha_{\rm OX} + 1.226 \times \alpha_{\rm OX}^2$, \citep{2010A&A...512A..34L, 2018MNRAS.480.1522L} we estimated the average bolometric luminosity of the AGN to be $\log \lbol \sim 45.23\ergs$ by plugging in the values of OM fluxes of \xmm{} observations, the values of $\rm L_{2-10\kev}$ and $\alpha_{\rm OX}$, where $\alpha_{\rm OX}$ is the power-law slope joining the $2 \kev$ and the $2500 \rm \AA$ flux. For this source, a black hole mass of $\sim 3.98\times 10^{7}\msol$, this implies an Eddington ratio of $\lambdaedd = 0.31$ which is consistent with previous estimates \citep{2016MNRAS.458.2454L, 2019MNRAS.484.5113R}. We detected the presence of a partially ionized absorbing material with a relatively high column density $\rm N_{H} =20.6^{+1.0}_{-1.0} \times 10^{22} cm ^{-2}$. The ionization parameter $\log\xi/ \xiunit =0.1^{+0.1}_{-0.1}$, and covering fraction found in our work, $0.21^{+0.01}_{-0.01}$, is consistent with \citet{2016MNRAS.458.2454L}. Current measurements of cutoff energies of the primary power-law component, range between 50 and $300 \kev$ \citep{2018MNRAS.473.3104T, 2019MNRAS.484.2735M, 2019A&A...629A..54U} for bright Seyfert 1 galaxies and require spectra extending well above $50\kev$ for better constraints. Using relativistic reflection model, we found a lower limit to the cutoff energy, $E_{c} > 276\kev$. Below we discuss the main results investigating the reflection features observed in this source.

\subsection{The soft excess}\label{subsec:variability}

We detected the presence of soft X-ray excess below $2\kev$ for all the \xmm{} observations of ESO~141--G055. We investigated the broadband multi-epoch data in detail to identify its possible origin. Recent studies show that blurred ionized reflection from the accretion disk can explain the origin of soft excess in several Seyfert 1 galaxies such as Mrk~509 \citep{2019ApJ...871...88G}, 1H~0323+342 \citep{2018MNRAS.479.2464G} and Mrk~478 \citep{2019MNRAS.489.5398W}. On the other hand, the intrinsic thermal Comptonisation model can describe the soft X-ray excess better in the sources such as Ark~120 \citep{2004MNRAS.351..193V, 2018A&A...609A..42P}, Mrk~530 \citep{2018MNRAS.478.4214E}, HE~1143-1810 \citep{2020A&A...634A..92U} and Zw~229.015 \citep{2019MNRAS.488.4831T}. We note that although these two models imply two very different interpretations of the observed spectra, it is not easy to distinguish them on statistical grounds alone. However, certain model parameters exhibited extreme values in certain fits which helped in ruling the models out e.g., in the sources HE~1143-1810 \citep{2020A&A...634A..92U} and Zw~229.015 \citep{2019MNRAS.488.4831T} the ionized disk reflection model was not found suitable to describe the soft excess because the iron abundance, the inclination angle and reflection fraction were too low compared to normal values expected in Seyfert 1 galaxies. On the other hand, in Mrk~478 \citep{2019MNRAS.489.5398W}, the flux variability between data sets favoured the ionized reflection over thermal Comptonisation model. 

In this study, with ESO~141--G055, We get similar fit statistics for both the models. The parameter values constrained for these models were in a range detected in typical Seyfert 1 galaxies (See Table \ref{Table:relxill} and Table \ref{Table:optxagnf}). The best-fit photon index $\Gamma$ of the relativistic reflection model {\it relxill} determines the shape of the primary continuum and ranges between $2.0 - 2.1$ for all observations. The ionization parameter of the reflecting disk is $\log \xi = 2.14^{+0.02}_{-0.02} \xiunit$, suggesting a low to a moderately ionized disk. The inner extent of the accretion disk ($\rm r_{in} =3.29^{+0.22}_{-0.24} \rm r_{g}$) and the high emissivity index of $7.25^{+0.24}_{-0.40}$ implies that major part of the soft excess has originated from a region very close to the central supermassive black hole. The inner radius and the black hole spin are degenerate in the {\it relxill} model and while fitting the soft excess with this model we obtained a maximally rotating black hole ($\rm a > 0.96$). The fit worsens statistically if we freeze the spin to zero ($\dc\sim 300)$. The maximally spinning black hole scenario has also been detected in previous studies in other Seyfert 1 galaxies \citep{2016MNRAS.456..554G, 2019ApJ...871...88G, 2018MNRAS.479.2464G, 2019MNRAS.489.5398W}. This high value of the spin is usually associated with a broad Fe K$_{\alpha}$ which we do not detect in any of our spectra. Hence, we presume that the high spin we obtained may be a necessity of the model to explain the soft excess, while in reality, the spin may not be that high. The reflection fraction (R) is defined as the ratio of photons from the hot corona striking the disk with that of the photons emitted from the hot corona escaping to infinity. We obtained a relatively low but well-constrained value of R ranging between $0.35^{+0.01}_{-0.01}$ to $0.54^{+0.01}_{-0.01}$. These results imply that despite the presence of a compact corona at or above the supermassive black hole, only a small amount of primary flux illuminates the accretion disk while most of it escaping to infinity. This also explains the low to moderate ionization state of the disk. These relatively lower values are consistent with the 2001 \xmm{} observation \citep{2003A&A...398..967G} and other Seyfert 1s as well, e.g., Zw~229.015, where these values are $\log \xi = 2.56^{+0.15}_{-0.15}\xiunit$ and $\rm R \sim 0.4$. The iron abundance of the material in the accretion disk and the inclination of the disk to the line of sight plays an important role in describing the reflected spectra of Seyfert 1 galaxies \citep{2006MNRAS.365.1067C, 2014ApJ...782...76G}. In our study, the iron abundance when made free was consistent with solar values ($1.04^{+0.17}_{-0.07}$) and hence was fixed to 1 for all observations. We were able to constrain the inclination angle to $10^{+2}_{-1}$ degree which is consistent with Seyfert type 1 classification criteria.

The {\it optxagnf} model provides a good fit to the soft excess. This model describes the soft excess as Comptonization of thermal disk photons by a warm and optically thick corona. This warm corona could be the upper layer of the accretion disk covering roughly $10-20 \rm r_{g}$ of the inner region \citep{2013A&A...549A..73P}. In the source ESO~141--G055, the best-fit temperature of the warm corona varies between  $\rm T \sim 0.1 - 0.3 \kev$ and we could put a lower limit on the opacity ($\tau >6$), implying an optically thick medium. The measured warm-corona radius ($\rm r_{\rm corona}$) is $9.09^{+0.01}_{-0.01} \rm r_{g}$. 
The black hole spin inferred by the {\it optxagnf} model is $\rm a > 0.98$ which is consistent with that of the {\it relxill} model. The fit statistically worsens (by $\dc \sim 250$) if we freeze the spin to zero. We think that the spin measurement is real but the absence of any broad Fe line prevents us from making any strong claim on the spin of the black hole. We note that similar results have been previously found by \citet{2019MNRAS.489.5398W} in other Seyfert 1s where narrow Fe line is detected. In addition, we required an OVII emission line in this fit which indicates ionized emission from the disk. Our analysis of the \xmm{} data with {\it optxagnf} model shows no significant variation in the Eddington ratio from 2001 to 2006 and is consistent with $\lambdaedd \sim 0.14$. This sub-Eddington rate is common among several Seyfert 1s. 
The accretion rate in {\it optxagnf} is also determined by the outer accretion flow beyond $\rm r_{corona}$ that contributes primarily in the optical/UV band. Due to the absence of simultaneous optical/UV data, we could not properly constrain the $\lambdaedd$ value in this spectral fit with the optxagnf model. The fraction of power below the coronal radius emitted in the hard comptonisation component is consistent with a high value of $\sim 0.9$. 
This implies around 90 per cent of the gravitational energy released below $9 \rm r_{g}$, is emitted as the primary X-ray emission with a photon index $\sim 2.1$ and the rest would contribute to the soft excess.

\subsection{The Fe line complex and the Compton hump}\label{subsec:contributor}

All the spectra of ESO~141--G055 show the presence of narrow Fe emission line at $\sim 6.4\kev$ and we did not find any signature of broad Fe line in any of the observations. The centroid energy of the Fe emission line indicates that the iron is neutral or in low ionization state, consistent with the ionization parameter of the disk reflector ($\log \xi \leq 2.14\xiunit $). The use of {\it diskline} model did not improve the fit statistics for either \nustar{} or the \xmm{} data. We note that although the broadband spectra of ESO~141--G055 favours the maximally rotating black hole scenario, we did not find any evidence of a broad Fe emission line. This may mean that either the broad Fe line is absent or blurred beyond detection.

\begin{figure*}
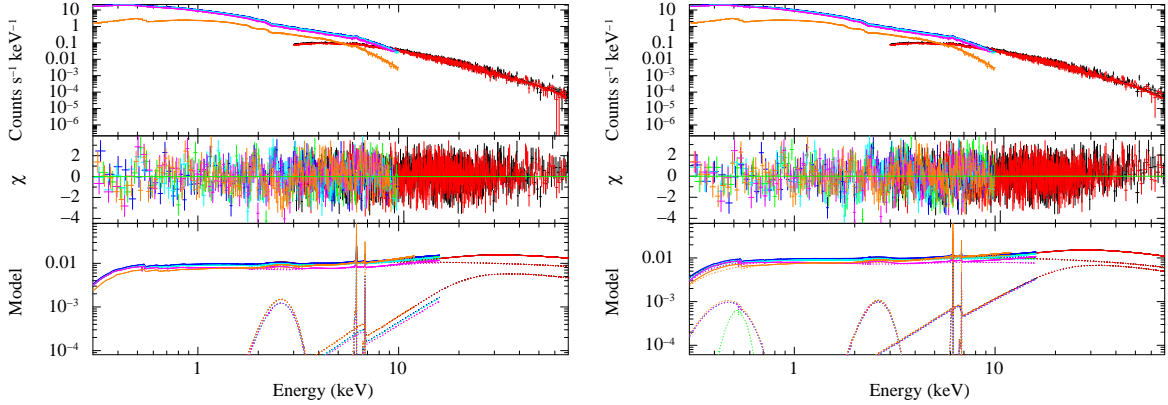

  \centering 

\hbox{
\includegraphics[width=5.5cm,angle=-90]{Best_fit_absorbed_relxill_mytorus_withspin_zxipcf_zgauss2.5keV.new.const.all.spec.eso141.ps}
\includegraphics[width=5.5cm,angle=-90]{Best_fit_absorbed_optxagnf_mytorus_with_spin_zxipcf.zgauss.all.spectra.eso141mg055.ps} 
}\caption{ {\it Left:} The $0.3-70.0\kev$ \nustar{} and \xmm{} spectra of the source ESO~141--G055 with the best-fitting reflection model and residuals. The relxill model describing simultaneously the soft X-ray excess, the broad Fe Kα emission line, and the relativistic reflection hump in the hard X-rays, the MYTorus model describing the narrow Fe K$_\alpha$ and Ni emission lines along with the Compton hump due to distant neutral reflection, are plotted in the lower panel. {\it Right:} Same for the Comptonization model optxagnf. The X-axis represents the observed frame energy. } \label{fig:phys}

\end{figure*}

The narrow Fe $\rm K_{\alpha}$ emission line detected at $\sim 6.4\kev$ has an equivalent width of $55-70 \ev$, typical of nearby Seyfert galaxies~\citep{1989MNRAS.236P..39N, 1989MNRAS.240..769P, 2011ApJ...727...19F, 2014MNRAS.441.3622R}. This narrow emission line is believed to arise when hard X-ray photons from the corona are reflected from high-column-density of neutral or lowly ionized material in the outer part of the disk and possibly the torus. The neutral reflection model {\it PEXRAV} and {\it MYTorus} can describe the Fe K emission lines consistently with the reflected continuum. Recent multiwavelength studies \citep{2011ApJ...738..147S, 2014A&A...567A.142R} of the narrow Fe $\rm K_{\alpha}$ revealed that the line originates from a distant region ($\sim 3 \times 10^{4} \rm r_{g}$). 
 \citet{2011ApJ...727...19F} found a strong correlation between the equivalent width of the neutral Fe $\rm K_{\alpha}$ emission line and the neutral absorption column density in the range $10^{23} - 10^{25}\rm cm^{-2}$. In ES0~141--G055, the {\it MYTorus} model yielded a good description of the narrow emission line and the Compton hump (peaking at $\sim 20 \kev$). The best-fit value of the reflection column density is found to be $\geq 7\times 10^{24} \rm cm^{-2}$, suggesting a Compton thick reflection. This result favours the fact that the neutral Fe $\rm K_{\alpha}$ line emission of ESO~141--G055 originates from the Compton-thick torus. We note that in ionized reflection model, the Compton hump has a contribution from distant neutral reflection from the torus. From Fig.~\ref{fig:phys}, we find that in the thermal comptonization model, neutral reflection alone can provide a good description of the Compton hump. With the current data quality in \nustar{}, we can not distinguish between these two models and hence unable to efficiently isolate the contributions of the ionized and neutral reflection components which describes the Compton hump. Further deep observations of these sources are required to accomplish that.

\section{Conclusions}\label{sec:conclusions}

We have extensively studied the broadband X-ray spectra of the local Seyfert 1 galaxy ESO~141--G055, investigating the presence and origin of the reflection features with a physically motivated set of models using \nustar{} and \xmm{} data. We list the main conclusions below.

\begin{itemize}

	\item{The X-ray spectra of ESO~141--G055 is typical of local Seyfert galaxies, with the powerlaw photon index varying between $\Gamma =2.0 - 2.1$, a prominent soft-excess, a narrow Fe line emission at $\sim 6.4\kev$, and a hard X-ray excess above $10 \kev$. Using reflection model {\it relxill}, we found a lower limit to the cutoff energy of the primary power-law component, $E_{c} > 276\kev$.}

	\item{ The soft X-ray excess is equally well described by both the models: the blurred reflection from ionized disk and the intrinsic thermal comptonization of disk photons. Both these models require a maximally spinning black hole to describe the soft excess. }

	\item{We found the presence of a narrow Fe emission line at $\sim 6.4\kev$ with an equivalent width of $ 55-70 \ev$, that is typical of nearby Seyfert galaxies. The best-fitting reflection column density of the {\it MYTorus} model is found to be $\geq 0.52 \times 10^{24} \rm cm^{-2}$ which indicates that the neutral Fe $\rm K_{\alpha}$ emission line of ESO~141--G055 has originated from neutral Compton thick reflector (possibly the torus).}

	\item {The cold, distant neutral reflection from torus is a major contributor to the Compton hump along with the ionized reflection. }

	\item{We detected a partially covering ionized absorption with ionization parameter $\log \xi/\xiunit $ = $0.1^{+0.1}_{-0.1}$, and a covering factor of $0.21^{+0.01}_{-0.01}$. The column density is found to be $\rm N_{H} =20.6^{+1.0}_{-1.0}\times 10^{22} \rm cm^{-2}$.}

\end{itemize}

\section{Acknowledgements}
The authors are grateful to the anonymous referee for insightful comments which improved the quality of the paper.
RG acknowledges the financial support from Visva-Bharati University and IUCAA visitor programme. 

\section{Data availability}

This research has made use of archival data of \nustar{} and \xmm{} observatories through the High Energy Astrophysics Science Archive Research Center Online Service, provided by the NASA Goddard Space Flight Center. 

\bibliographystyle{mnras}
\bibliography{mybib}

\begin{table*}

{\footnotesize
\centering
  \caption{The X-ray observations of ESO~141--G055. \label{Table:obs}}
  \begin{tabular}{cccccccc} \hline\hline 

X-ray		& observation	&Short	&Date of obs	& Net exposure	\\
Satellite	&id		&id	&		&	\\ \hline 

\xmm{}          &{0101040501}   &xmm01  &09-10-2001     & $55\ks$       \\
		&{0503750301}	&xmm02	&09-10-2007	& $31\ks$	\\
		&{0503750401}	&xmm03	&11-10-2007	& $27\ks$	\\
		&{0503750501}	&xmm04	&12-10-2007	& $76\ks$	\\
		&{0503750101}	&xmm05	&30-10-2007	& $80\ks$	\\

\nustar{}	&60201042002	&	&15-07-2016	&$93\ks$	\\

\hline 
\end{tabular}  


}
\end{table*}



\begin{table*}

\centering

  \caption{The best fit parameters of the baseline phenomenological models for the \nustar{} and \xmm{} observations of ESO~141--G055.  \label{Table:pheno}}
{\renewcommand{\arraystretch}{1.5}
\setlength{\tabcolsep}{1.5pt}
\begin{tabular}{cccccccc} \hline\hline
	
Models 		& Parameter 				& NuSTAR 		&xmm01	     & xmm02 		   & xmm03 		& xmm04 		 & xmm05  \\ \hline 

Gal. abs.  	& $\nh \,(\times 10^{20}\, \cmsqi)$ 	& $ 4.94$ (f)     	&$ 4.94$ (f) & $ 4.94$ (f)     	   & $ 4.94$ (f)     	& $ 4.94$ (f)     	 & $ 4.94$ (f) \\

zxipcf   	&   $\nh$ ($\times 10^{22}\cmsqi$)      & ---   	 	&$0.39^{+0.14}_{-0.12}$& $0.84^{+0.29}_{-0.35}$   & $0.64^{+0.21}_{-0.47}$ 	& $0.29^{+0.14}_{-0.18}$ & $0.38^{+0.34}_{-0.32}$  \\
                &   $log\xi$                            & ---   	 	&$-0.5^{+0.2}_{-0.8}$  & $-0.4^{+0.6}_{-0.3}$    & $-0.5^{+0.2}_{-0.6}$ 	& $-0.4^{+0.2}_{-0.1}$	 & $-0.2^{+0.3}_{-0.2}$  \\
                &   $f$                                 & ---   	 	&$0.5^{+0.1}_{-0.1}$   & $0.6^{+0.1}_{-0.3}$ 	   & $0.5^{+0.2}_{-0.2}$ 	& $0.5^{+0.1}_{-0.2}$	 & $0.3^{+0.2}_{-0.1}$ \\

 powerlaw 	& $\Gamma$         			& $1.93^{+0.03}_{-0.03}$ &$1.87^{+0.05}_{-0.03}$& $1.83^{+0.07}_{-0.09}$  & $1.95^{+0.14}_{-0.14}$ & $2.16^{+0.06}_{-0.06}$  & $1.79^{+0.03}_{-0.04}$  \\
        	& norm ($10^{-3}$) 			& $8.56^{+0.04}_{-0.03}$ &$7.93^{+0.67}_{-0.37}$& $7.81^{+1.09}_{-1.19}$  & $7.64^{+1.21}_{-1.15}$ & $8.92^{+0.45}_{-0.53}$  & $6.95^{+0.40}_{-0.47}$  \\

diskbb  	& $T_{in}$ (keV)   			& ---                   &$0.04^{+0.02}_{-0.01}$ & $0.06^{+0.01}_{-0.01}$   & $0.09^{+0.03}_{-0.03}$ & $0.04^{+0.02}_{-0.02}$& $0.10^{+0.02}_{-0.03}$    \\
          	& norm ($10^{4}$) 			& ---                   &$9.27^{+2.85}_{-1.44}$ & $9.86^{+5.51}_{-1.63}$   & $1.80^{+3.73}_{-1.31}$ & $8.91^{+4.29}_{-2.26}$& $0.89^{+0.81}_{-0.31}$    \\

diskbb  	& $T_{in}$ (keV)   			& ---                   &$0.19^{+0.03}_{-0.01}$ & $0.24^{+0.03}_{-0.03}$   & $0.26^{+0.02}_{-0.02}$ & $0.17^{+0.03}_{-0.01}$& $0.30^{+0.02}_{-0.03}$    \\
          	& norm ($10^{3}$) 			& ---                   &$1.01^{+0.49}_{-0.67}$& $0.79^{+1.13}_{-6.56}$   & $0.34^{+0.55}_{-0.28}$ & $1.27^{+0.56}_{-0.71}$& $0.78^{+0.13}_{-0.07}$    \\

Gaussian	&E($\kev$)        			& $6.39^{+0.09}_{-0.09}$ &$6.62^{+0.15}_{-0.15}$& $6.33^{+0.07}_{-0.08}$  & $6.33^{+0.05}_{-0.05}$ & $6.35^{+0.05}_{-0.05}$& $6.43^{+0.07}_{-0.07}$ \\
		&$\sigma$($\kev$)        		& $0.31^{+0.11}_{-0.12}$ &$0.35^{+0.20}_{-0.11}$& $<0.18$		   & $<0.14$                & $<0.10$		    & $0.12^{+0.09}_{-0.09}$ \\
		&norm ($10^{-5}$) 			& $3.38^{+0.70}_{-0.71}$ &$4.51^{+1.60}_{-1.96}$& $1.54^{+0.76}_{-0.70}$  & $1.50^{+0.72}_{-0.64}$ & $1.34^{+0.60}_{-0.51}$& $1.81^{+0.61}_{-0.47}$ \\

		&$\rm ^A$$\rm \dc/dof$			&$113/3$		 &$28/3$    &$35/3$		   &$50/3$		    &$57/3$		 &$62/3$	\\

Pexrav $^{B}$           & R              		& $-0.39^{+0.08}_{-0.16}$ &$-0.2^{*}$ & $-1.9^{*}$          & $-1.07^{+0.85}_{-1.10}$ & $-1.68^{+0.57}_{-0.65}$    &  $>-0.22$                \\
			& Incl             		& $10^{*}$            	  &$10^{*}$ & $10^{*}$              & $10^{*}$ 		        & $10^{*}$               & $10^{*}$                \\
			&$\rm ^A$$\rm \dc/dof$          & $97/2$                  & --       & ---                 & $27/2$ 		        & $13/2$      		 &   ---                   \\

Gaussian  	&  EqW (eV)  				 & 98                    & 76       &52                     & 55 	    &  56            & 73                     \\\hline 
 
$\cd$     &                  				& $ 1334/1284$            &$ 158/161 $ & $ 142/141 $            & $ 166/137 $            & $ 171/134 $    & $ 201/147 $             \\\hline
\end{tabular}  

{$\rm ^A$ The $\dc$ improvement in statistics upon addition of the corresponding discrete component.}\\
{$\rm ^B$ The model {\it pexrav} was used only for \nustar{} observation as it had broad band spectra necessary for constraining the parameters. The values quoted for the \xmm{} observations are from the simultaneous fit of all the data sets.\\
 (*) indicates parameters are not constrained}\\
}

\end{table*}


\begin{table*}
\footnotesize
\centering
  \caption{Best fit parameters for observations of ESO~141--G055 with the first set of physical models. In XSPEC, the models read as {\it(constant $\times$ tbabs$\times $zxipcf$\times$(relxill + MYTorus))}. \label{Table:relxill}}

  \begin{tabular}{llllllll} \hline
Component  & parameter                	     & \nustar{}     	      &xmm01       & xmm02 		& xmm03 		& xmm04 		& xmm05  \\\hline

Gal. abs.  & $\nh (10^{20} cm^{-2})$  	     & $ 4.94$ (f)     	      &$ 4.94$ (f) & $ 4.94$ (f)     & $ 4.94$ (f)     	& $ 4.94$ (f)           & $ 4.94$ (f) \\

zxipcf    &   $\nh$ ($\times 10^{22}\cmsqi$) & $20.6$ (t)  &$20.6$ (t) &$20.6^{+1.0}_{-1.0}$        &$20.6$(t)              & $20.6$(t)             & $20.6$(t) \\

          &  $\log \xi(\ergs)$                & $0.1$(t)    &$0.1$(t) & $0.1^{+0.1}_{-0.1}$        &$0.1$(t)               & $0.1$(t)              & $0.1$(t) \\

          &  $Cvr_{\rm frac}$                & $0.21$(t)   &$0.21$(t) & $0.21^{+0.01}_{-0.01}$     & $0.21$(t)             & $0.21$(t)             & $0.21$(t)  \\\hline

{\it relxill }  &  $A_{Fe}$                  & $1$(f)  	   & $1$(f)   & $1$(f)                     & $1$(f)    	        & $1$(f)                & $1$(f)  \\
 
           	&  $\log\xi (\xiunit)$       & $2.14^{+0.02}_{-0.02}$  &$2.14$ (t) & $2.14$ (t)      & $2.14$ (t)            & $2.14$ (t)            & $2.14$ (t)  \\ 

           	& $ \Gamma $               & $2.03^{+0.01}_{-0.01}$   &$1.97^{+0.01}_{-0.01}$ & $2.07^{+0.01}_{-0.01}$&$2.08^{+0.01}_{-0.01}$ &$2.14^{+0.01}_{-0.01}$  &$2.07^{+0.01}_{-0.01}$\\

		& $\rm E_{cut}(\kev)$      & $>276$                   &$301$(t) 	            & $301$(t)         & $301$ (t)          & $301$(t)             & $301$(t)\\

           	&  $n_{rel}(10^{-5})^a$    & $18.11^{+0.09}_{-0.05}$  &$14.48^{+0.21}_{-0.22}$ &$21.01^{+0.06}_{-2.89}$&$19.24^{+0.05}_{-2.65}$ &$15.62^{+0.05}_{-0.05}$ &$19.66^{+0.02}_{-2.70}$ \\
	   
           	&   $ q1$                  & $7.25^{+0.11}_{-0.12}$   &$7.25^{+0.24}_{-0.40}$       & $7.25$(t)          & $7.25$(t)          &$7.25$(t)              &$7.25$(t) \\
 
		&   $ a$                   & $>0.96$    	      &$0.99$(t) & $0.99$(t)         & $0.99$ (t)          & $0.99$(t)             & $0.99$(t)\\

           	&   $R(\rm refl frac) $    & $0.10^{+0.01}_{-0.01}$   &$0.47^{+0.05}_{-0.05}$ & $0.35^{+0.01}_{-0.01}$& $0.35^{+0.01}_{-0.01}$ & $0.54^{+0.01}_{-0.01}$   & $0.35^{+0.01}_{-0.01}$  \\

           	&   $ R_{in}(r_{g})$       & $3.06^{+0.03}_{-0.03}$   &$3.29^{+0.22}_{-0.24}$ & $3.06$(t)       & $3.06$(t)             & $3.06$(t)             & $3.06$(t)  \\

		&   $ R_{br}(r_{g})$       & $7.2^{+0.3}_{-0.3}$     &$7.1^{+0.4}_{-0.4}$ & $7.1$(t)       & $7.1$(t)             & $7.1$(t)             & $7.1$(t)  \\

           	&   $ R_{out}(r_{g})$      & $400$ (f)      	      &$400$(f) & $400$(f)         & $400$(f)              & $400$(f)              & $400$(f)    \\

           	&   $i(\rm degree) $       & $10^{+2}_{-1}$	      &$10$(t)  & $10$(t)          & $10 $(t)              & $10$(t)               & $10$(t) \\ \hline

{\it MYTorusL}  &   $i(\rm degree) $       & $>45$                    &$ 57$(t)  & $ 57$(t)         & $ 57$(t)              & $57$(t)               & $57$(t) \\

		&  norm ($10^{-3}$)        & $17.95^{+2.50}_{-2.03}$  &$17.95$(t) & $17.95$(t)      & $17.95$(t)            &$17.95$(t)             & $17.95$(t) \\

{\it MYTorusS } &  NH($10^{24}\rm cm^{-2}$)& $>7.1$                   &$10$(t) & $10$(t)            &$10$(t)                   &$10$(t)                & $10$(t)\\

		&  norm ($10^{-3}$)        & $14.20^{+2.74}_{-0.68}$  &$14.20$(t) & $14.20$(t)      &$14.20$(t)             &$14.20$(t)             &$14.20$(t)\\\hline
		
	   	& $\cd $                   & $1342/1271$      	      &$191/168$  & $165/134$       &$185/135$              &  $189/129$  & $242/159$ \\\hline 
\end{tabular} \\ 
Notes: (f) indicates a frozen parameter. (t) indicates a tied parameter between observations. \\
	(a) $n_{rel}$ reperesent normalization for the model {\it relxill}\\
\end{table*}

\begin{table*}

\footnotesize
\centering
  \caption{Best fit parameters for observations of ESO~141--G055 with the second set of physical models. In XSPEC, the models read as {\it(constant $\times$ tbabs$\times $zxipcf$\times$(optxagnf+ MYTorus))}.\label{Table:optxagnf}}

  \begin{tabular}{llllllll} \hline

Component  & parameter                	     & \nustar{}     	      & xmm01  & xmm02 		& xmm03 		& xmm04 		& xmm05  \\\hline


Gal. abs.  & $\nh (10^{20} cm^{-2})$  	     & $ 4.94$ (f)     	      &$ 4.94$ (f) & $ 4.94$ (f)     & $ 4.94$ (f)     	& $ 4.94$ (f)           & $ 4.94$ (f) \\

zxipcf    &   $\nh$ ($\times 10^{22}\cmsqi$) & $13.8$(t)       &$13.8$(t) &  $13.8^{+2.1}_{-2.0}$  &$13.8$(t)              & $13.8$(t)             & $13.8$(t) \\
          &  $\log \xi(\ergs)$                & $-0.2$(t)       &$-0.2$(t) & $-0.2^{+0.2}_{-0.2}$   &$-0.2$(t)              & $-0.2$(t)             & $-0.2$(t) \\
          &  $Cvr_{\rm frac}$                 & $0.28$(t)       &$0.28$(t) & $0.28^{+0.02}_{-0.02}$ & $0.28$(t)             & $0.28$(t)             & $0.28$(t)  \\\hline

optxagnf   & $ \rm M_{BH}^a$        &$3.98$(f)                &$3.98$(f) & $3.98$(f)               &$3.98$(f)               &$3.98$(f)              &$3.98$(f)\\
           & $d{\rm~(Mpc}) $        &$163$(f)                 &$163$(f) & $163$(f)                &$163$(f)                &$163$(f)               &$163$(f)\\
           &  $(\frac{L}{L_{E}})$   &$0.42^{+0.01}_{-0.06}$   &$0.13^{+0.04}_{-0.02}$ & $0.15^{+0.03}_{-0.02}$  &$0.14^{+0.04}_{-0.01}$  &$0.12^{+0.03}_{-0.02}$ &$0.14^{+0.04}_{-0.03}$\\ 
           &  $ kT_{e} (\kev)$      &$0.08$(t)                &$0.26^{+0.02}_{-0.02}$ & $0.08^{+0.01}_{-0.01}$  &$0.08$(t)               &$0.08$(t)              &$0.08$(t) \\ 
           &  $ \tau $              &$15$(t)                  &$15$(t) & $>6$                    &$15$(t)                 &$15$(t)                &$15$(t)\\
           &  $ r_{\rm cor}(r_{g})$ &$2.9^{+0.2}_{-0.2}$      &$>4.1$ & $9.09^{+0.01}_{-0.01}$  &$9.09$(t)               &$9.09$(t)              &$9.09$(t)\\
           &  $ a $                 &$0.99$(t)                &$0.99$(t) & $>0.98$              &$ 0.99$(f)              &$ 0.99$(f)             &$0.99$(f) \\
           &  $ f_{pl}$             &$0.90^{+0.04}_{-0.03}$   &$0.95^{+0.01}_{-0.01}$ & $0.95^{+0.01}_{-0.01}$  &$0.96^{+0.01}_{-0.01}$  &$0.93^{+0.01}_{-0.01}$ &$0.96^{+0.01}_{-0.01}$ \\
           &  $ \Gamma $            &$2.11^{+0.01}_{-0.01}$   &$1.93^{+0.02}_{-0.02}$ &$2.08^{+0.01}_{-0.01}$  &$2.09^{+0.01}_{-0.01}$  &$2.13^{+0.01}_{-0.01}$ &$2.08^{+0.01}_{-0.01}$ \\\hline

{\it MYTorusL}  &   $i(\rm degree) $       & $>51$                    &$ 60$(t) & $ 60$(t)        & $ 60$(t)              & $60$(t)               & $60$(t) \\
		&  norm ($10^{-3}$)        & $18.15^{+0.02}_{-0.10}$  &$18.15$(t) & $18.15$(t)      & $18.15$(t)            &$18.15$(t)             & $18.15$(t) \\

{\it MYTorusS } &  NH($10^{24}\rm cm^{-2}$)& $10.0$(*)         	      &$10.0$(t) & $10.0$(t)       &$10.0$(t)              &$10.0$(t)              & $10.0$(t)\\
		&  norm ($10^{-3}$)        & $30.74^{+0.01}_{-0.01}$  &$30.74$(t) & $30.74$(t)      &$30.74$(t)             &$30.74$(t)             & $30.74$(t) \\\hline
		
	   	& $\cd $                   & $1349/1277$  & $193/169 $   &$172/133 $         & $193/134 $    &$ 171/128$   &$258/160 $ \\\hline 
\end{tabular} \\ 
Notes: (f) indicates a frozen parameter. (t) indicates a tied parameter between different observation.\\(*) indicates parameters are not constrained. \\(a):in units of $10^7\rm M\odot$;\\
\end{table*}




\clearpage

\end{document}